# Realization of Semiconducting Layered Multiferroic Heterojunctions via Asymmetrical Magnetoelectric Coupling

Baishun Yang,[1,2] Bin Shao,[1,2] Jianfeng Wang,[1] ChiYung Yam,[1,2] Shengbai Zhang,[3] Bing Huang[1,4*]

[1]*Beijing Computational Science Research Center, Beijing 100193, China*
[2]*Shenzhen JL Computational Science and Applied Research Institute, Shenzhen, 518109, China*
[3]*Department of Physics, Applied Physics and Astronomy, Rensselaer Polytechnic Institute, Troy, New York 12180, USA*
[4]*Department of Physics, Beijing Normal University, Beijing, 100875, China.*

Email: bing.huang@csrc.ac.cn

**Two-dimensional (2D) semiconducting multiferroics that can effectively couple magnetic and polarization (*P*) orders have great interest for both fundamental research and technological applications in nanoscale, which are, however, rare in nature. In this study, we propose a general mechanism to realize semiconducting 2D multiferroics via vdW heterojunction engineering, as demonstrated in a typical heterostructure consisting of magnetic bilayer $CrI_3$ (*bi*-$CrI_3$) and ferroelectric monolayer $In_2Se_3$. Interestingly, the novel indirect orbital coupling between Se 4*p* and Cr 3*d* orbitals, intermediated by the interfacial I 5*p* orbitals, are switchable in the opposite *P* configurations, resulting in an unexpected mechanism of strong asymmetrical magnetoelectric coupling. Therefore, along with the noticeable ferroelectric energy barrier induced by $In_2Se_3$, the realization of opposite magnetic orders in opposite *P* configurations can eventually result in the novel multiferroicity in *bi*-$CrI_3$/$In_2Se_3$. Finally, we demonstrate that our mechanism can generally be applied to design other vdW multiferroics even with tunable layer thickness.**

Multiferroic materials that can simultaneously process more than one primary ferroic order in the same phase, including ferromagnetic (FM), ferroelectric (FE), or ferroelastic order [1-3], have attracted great attentions for both scientific interest and technological applications [4,5]. In general, there are two types of multiferroics: type-I has a weak magnetoelectric (ME) coupling because of the different FM and FE origins [6]; type-II has a strong ME coupling, but the electric polarization ($P$) originated from the spin-orbital coupling is usually weak [7]. Meanwhile, the spontaneous $P$ could be suppressed when a FE system is below its critical thickness, hindering their opportunities in nanoscale. Compared to the bulk FE systems, the two-dimensional (2D) layered FE systems with reduced surface energies may eliminate the intrinsic depolarization field in the monolayer (ML) limit [8,9]. Motivated by the discovery of 2D FE and magnetic systems [10-23], it is interesting to further explore the 2D multiferroics. Indeed, a few 2D systems, *e.g.*, transition-metal (TM) halides [24], double metal trihalides [25], and TM phosphorus chalcogenides [26,27] are predicted to be 2D single-phase multiferroics, however, in practice their artificially designed structures have so far prevented their realization (problem i).

Recently, the van de Waals (vdW) heterojunction engineering, *via* the stacking of layered systems with different properties, has provided a new opportunity to realize exotic physical properties and functionalities beyond those of their individual components [14,28,29]. Naturally, it is desirable to explore the feasibility of realizing vdW double-phase layered multiferroics via the ME coupling of 2D magnetic and FE systems. Actually, several attempts have been made, *e.g.*, for $CrI_3/Sc_2CO_2$ [30,31], $FeI_2/In_2Se_3$ [32] and $MnCl_3/CuInP_2S_6$ [33]. In these systems, an interface-charge-transfer-induced semiconductor-to-metal transition is required for achieving the magnetic ground-state switching under different $P$ configurations, which, however, fundamentally eliminates their intrinsic semiconducting properties (problem ii).

In this study, we propose a strategy of vdW heterojunction engineering of existing 2D systems to realize 2D semiconducting double-phase multiferroics, which can fundamentally overcome the above problems i and ii. Taking the prototype bilayer $CrI_3$ (*bi*-$CrI_3$)/$In_2Se_3$ system as an example, the inclusion of ML FE $In_2Se_3$ induces a large FE energy barrier ($E_b$) into the system, which can effectively separate the different magnetic ground states, *i.e.*, antiferromagnetic (AFM) and FM states, into opposite $P$ configurations, *i.e.*, $P\downarrow$ and $P\uparrow$. Importantly, the novel indirect coupling between the Se $4p$ and Cr $3d$ orbitals, mediated by the interfacial I $5p$ orbitals, are easily switchable in the opposite $P$ configurations to result in a novel mechanism of strong asymmetrical ME coupling. In other words, the existence of the AFM-$P\downarrow$ and FM-$P\uparrow$ ground states, together with the large $E_b$, eventually results in the semiconducting multiferroic *bi*-$CrI_3$/$In_2Se_3$. Finally, we demonstrate that the mechanism is not system specific but can be generally applied to design other vdW multiferroics, even with tunable layer thickness.

The concept of the proposed vdW multiferroic heterojunction is illustrated in **Fig. 1**. Basically, it includes two components: a vdW semiconducting bilayer magnet (*bi*-magnet) (**Fig. 1a**) and a FE ML (**Fig. 1b**). As shown in **Fig. 1a**, in the *bi*-magnet, its interlayer magnetic order could be either AFM (↑↓, left panel) or FM (↑↑, right panel). Without loss of generality, we assume that the AFM order is the ground state, which is lower in energy than the FM state by $\Delta E$. Interestingly, it is known that an external electric field ($E_{ef}$) along the out-of-plane (OOP) direction can induce a potential energy difference $\Delta V$ between the A and B layers in the *bi*-magnet, as shown in **Fig. 1a**. This lifts the AFM band degeneracy and consequently gives rise to an AFM-to-FM transition. When the $E_{ext}$ is withdraw, the FM phase may turn back to the AFM phase upon a small thermal perturbation, because the energy barrier $E_b$ between the AFM and FM phases, originated from the magnetic exchange coupling ($E_{xc}$) and the magnetic anisotropy energy (MAE), is usually tiny (only several meV). On the other hand, in a typical FE system (**Fig. 1b**), there is a relatively large FE $E_b$ (on the order of tens or hundreds of meV) between two equivalent phases with opposite $P$ directions. Therefore, a sufficiently large $E_{ef}$ is required to switch between the $P\uparrow$ and $P\downarrow$ states.

Our idea is to induce a large $E_b$ from FE ML to the *bi*-magnet to create a sizable barrier between the AFM and FM states, whereby realizing multiferroicity in the vdW heterojunction. As shown in **Fig. 1c**, upon attaching the *bi*-magnet on a FE layer, we expect that there will be both the AFM and FM states for each $P$ configuration. To realize the multiferroic function, the key is to achieve opposite magnetic ground-states in the $P\uparrow$ and $P\downarrow$ configurations, *e.g.*, AFM-$P\downarrow$ and FM-$P\uparrow$ (**Fig. 1c**). As importantly, by having a large FE $E_b$, it will not be overly easy to switch between these two states. In the following, we will demonstrate that such an asymmetric magnetic ground-states in the $P\uparrow$ and $P\downarrow$ configurations could be achieved by a novel mechanism of asymmetric ME coupling under the opposite $P$ configurations.

Below, the above design principle will be demonstrated for the prototypical *bi*-CrI$_3$/In$_2$Se$_3$ system, based on extensive first-principles calculations (See computational details in Supplemental Material [34]). The intralayer $E_{xc}$ of ML CrI$_3$ is always FM, while the interlayer $E_{xc}$ can be either FM or AFM, depending on stacking sequence [40,41]. A weak interlayer $E_{xc}$ in *bi*-CrI$_3$ also provides the opportunity for an AFM-FM transition via an electrostatic gating [42-44] or by applying a hydrostatic pressure [45,46]. On the other hand, the ML In$_2$Se$_3$, consisting of [Se-In-Se-In-Se] quintuple layers, has been confirmed to be a good FE system with a spontaneous $P$ along the OOP direction [15,16]. As shown in **Fig. 2a**, the $P$ order of In$_2$Se$_3$ depends on the movement of the middle Se-2 layer, *i.e.*, the down-shift and up-shift of the Se-2 layer will correspond to a $P\uparrow$ and $P\downarrow$ configuration, respectively. In our calculation, a 1×1 *bi*-CrI$_3$ and a $\sqrt{3}\times\sqrt{3}$ ML In$_2$Se$_3$ have been used to construct the vdW *bi*-CrI$_3$/In$_2$Se$_3$ supercell, due to their good lattice matching.

**Fig. 2b** shows $\Delta E$ between the AFM and FM states in *bi*-CrI$_3$, as a result sliding between the top and bottom layers of *bi*-CrI$_3$. Along the sliding path, the magnetic ground-state can be either FM (for 0<$x$<0.21 and $x$>0.35) or AFM (for 0.21<$x$<0.35) (see the red curve), which agree with previous calculations [41]. Importantly, this $\Delta E$ curve can be down (up) shifted towards a lower (higher) energy position in the bi-CrI$_3$/In$_2$Se$_3$ under the $P\uparrow$ ($P\downarrow$) configuration, indicating the existence of an ME coupling between bi-CrI$_3$ and In$_2$Se$_3$. Impressively, in the range of 0.16<$x$<0.37, the bi-CrI$_3$/In$_2$Se$_3$ can exhibit the FM (blue curve) and AFM (green curve) ground-states in the $P\uparrow$ and $P\downarrow$ configurations, respectively. For multiferroics, it is necessary to estimate $E_b$ between the FM-$P\uparrow$ and AFM-$P\downarrow$ configurations. Without In$_2$Se$_3$, this $E_b$ is as small as ~0.65 meV/Cr, which is induced by the intrinsic $E_{xc}$ and MAE (**Fig. S1** [34]). When forming the vdW junction with In$_2$Se$_3$, taking the typical HT-phase bi-CrI$_3$ as an example (see **Fig. 2c**), a double-well potential energy surface appears, which is in line with our design principle in **Fig. 1c**. Importantly, the calculated $E_b$ between the FM-$P\uparrow$ and AFM-$P\downarrow$ configurations is as large as ~459.0 meV/Cr, which is sufficiently high to stabilize the FM and AFM states in different $P$ energy wells. The $\Delta E$ between the FM-$P\uparrow$ and AFM-$P\downarrow$ states is ~6.9 meV/Cr. The opposite magnetic ground-states in the $P\downarrow$ and $P\uparrow$ configurations, together with the large $E_b$, is a strong indication that multiferroicity should be realized in this bi-CrI$_3$/In$_2$Se$_3$ system.

It is desirable to understand the respective physical origins of the FM and AFM ground-states in the $P\uparrow$ and $P\downarrow$ configurations, which serves as the key for achieving multiferroicity in bi-CrI$_3$/In$_2$Se$_3$. Generally speaking, these systems are all semiconducting regardless the order of $P$ (see **Fig. S2** [34]). In other words, charge transfer between In$_2$Se$_3$ and bi-CrI$_3$ is negligible, which is fundamentally different from other proposed 2D vdW multiferroic systems (overcome problem ii), where charge-transfer-induced semiconductor-to-metal transition is the key factor to achieve the AFM-FM transition [30-33]. Due to the $C_{3v}$ crystal symmetry, the I 5$p$ (Se 4$p$) orbitals split into doublet $e_1$ ($e_2$) and singlet $a_1$ ($a_2$) states, which are found in bi-CrI$_3$/In$_2$Se$_3$($P\uparrow$) to be mostly located in different energies with small overlaps (see shaded areas in **Fig. 3a**). **Fig. 3b** shows the orbital-projected band structure, which reveals that orbital coupling (e.g., in the range of -2~0 eV) between the occupied I-1 5$p$ and Se-1 4$p$ states is negligible (see wavefunction in **Fig. S3a** [34]). Meanwhile, the $P\uparrow$ configuration of In$_2$Se$_3$ can lead to a positive $E_{ef}$ across bi-CrI$_3$, which can effectively lower the potential energy of the Cr-1 3$d$ states by $\Delta V$ with respect to that of the Cr-2 3$d$ states (see **Fig. 3c**). Importantly, this $\Delta V$ will increase the strength of virtual hopping between empty Cr-1 $e_g$ and occupied Cr-2 $t_{2g}$ orbitals in the same spin channel (**Fig. 3a**) to result in the favored FM order in bi-CrI$_3$/In$_2$Se$_3$($P\uparrow$).

In contrast, for bi-CrI$_3$/In$_2$Se$_3$($P\downarrow$), the closer distance between Se-2 and Se-1 layers increases their orbital coupling strength, pushing the (antibonding) Se-1 states to higher energy positions. Now, the energy levels of interfacial I-1 $a_1$ ($e_1$) and Se-1 $a_2$ ($e_2$) states can have significantly larger overlaps, as indicated by the shaded areas in **Fig. 3d** [34]. Therefore, the selective orbital coupling between I-1 5$p$ and Se-1 4$p$ states is largely turned on. Meanwhile, since the I 5$p$ and Cr 3$d$ orbitals in bi-CrI$_3$ can strongly couple forming covalent bonds, the Se-1 4$p$ orbitals can also indirectly couple with Cr-1 3$d$ orbitals, intermediated by the I-1 5$p$ orbital. **Fig. 3e** [34] shows that there are significant triple orbital resonances (*e.g.*, in the range of -2~0 eV) among the occupied Se-1 $4p$, I-1 $5p$ and Cr-1 $3d$ states in the valence band of *bi*-CrI$_3$/In$_2$Se$_3$($P\downarrow$) (see wavefunction in **Fig. S3b**). Therefore, this novel multiply orbital coupling will significantly lower the orbital levels of Cr-1 compared to that of Cr-2. However, the In$_2$Se$_3$ $P\downarrow$ configuration induced negative $E_{ef}$ would simultaneously lower the potential energy of Cr-2 $3d$ states with respect to that of Cr-1 ones, in contrast to the positive $E_{ef}$ effect in *bi*-CrI$_3$/In$_2$Se$_3$($P\uparrow$). Unexpectedly, the competition between the novel indirect orbital coupling that lowers Cr-1 orbital levels and the negative $E_{ef}$ effect that lowers Cr-2 orbital levels can eventually result in a rather small $\Delta V$, as shown in **Fig. 3f**. Therefore, different from *bi*-CrI$_3$/In$_2$Se$_3$($P\uparrow$), the virtual hopping strength between empty Cr-1 $e_g$ and occupied Cr-2 $t_{2g}$ orbitals cannot be noticeably enhanced (**Fig. 3d**), which prevents the AFM-to-FM transition. The above mechanism for the HT-phase *bi*-CrI$_3$/In$_2$Se$_3$ can be readily applied to other *bi*-CrI$_3$/In$_2$Se$_3$ systems with 0.16<$x$<0.37 (**Fig. 2b**) to understand the existence of $P\uparrow$-FM and $P\downarrow$-AFM ground-states. Although the $E_{xc}$ strength is changed in the ranges of 0<$x$<0.16 and $x$>0.37, the indirect orbital coupling mechanism is not sufficiently strong to overcome the $\Delta E$ between AFM and FM states, hindering the realization of FM-AFM switching.

Note that this asymmetrical ME coupling can also induce different magnetic ground-states in a trilayer CrI$_3$ (*tri*-CrI$_3$) system. Taking the HT-phase CrI$_3$ again as an example, without the In$_2$Se$_3$ layer, the ground-state of *tri*-CrI$_3$ should exhibit an interlayer ↑↓↑ spin order, which agrees well with experiments [47] and is lower in energy than other spin configurations (**Fig. S4** [34]). With the inclusion of In$_2$Se$_3$, on the other hand, there will be an asymmetric distribution of the ground-state spin orders between the $P\uparrow$ and $P\downarrow$ configurations, as shown in **Fig. 4**. In the $P\downarrow$ state, the ground-state spin order maintains to be ↑↓↑. However, in the $P\uparrow$ state, the ground-state spin order is dramatically changed to become ↓↓↑. From the study of *bi*-CrI$_3$/In$_2$Se$_3$, we know that *bi*-CrI$_3$ favors the AFM (FM) order in the $P\downarrow$ ($P\uparrow$) configuration. Therefore, the favorable spin order of the bottom two CrI$_3$ layers in *tri*-CrI$_3$/In$_2$Se$_3$ (next to In$_2$Se$_3$) follows the same trend as in *bi*-CrI$_3$/In$_2$Se$_3$, which gives rise to the ↑↓↑-$P\downarrow$ and ↓↓↑-$P\uparrow$ ground states. Generally speaking, both *tri*-CrI$_3$ and *tri*-CrI$_3$/In$_2$Se$_3$ (**Fig. S5** [34]) systems favor low-spin configurations. Together with the large $E_b$ ~551.9 meV/Cr, the *tri*-CrI$_3$/In$_2$Se$_3$ system can also sustain multiferroicity with tunable interlayer spin orders, even when the total magnetic moments under different $P$ configurations are unchanged. Furthermore, it is reasonable to expect that such exotic multiferroicity can also be realized in other multiple-layer CrI$_3$/In$_2$Se$_3$ systems with different layer thickness.

Before closing, we would like to emphasize that our mechanism here also generally applies to design other layered double-phase multiferroic heterojunctions. As shown in **Fig. S6** [34], a similar AFM-$P\uparrow$ and FM-$P\downarrow$ states can also be realized in the *R*-type *bi*-CrBr$_3$/In$_2$Se$_3$ system, as a result of a similar strong asymmetrical ME coupling mechanism under different $P$ configurations. Given that the vdW magnet, *e.g.*, CrI$_3$ [11] and CrBr$_3$ [48], and FE ML, *e.g.*, In$_2$Se$_3$ [15] and CuInP$_2$S$_6$ [20], have now been widely synthesized, our concept here can thus be readily tested by experiments.

In summary, we have proposed a generalized strategy in achieving semiconducting multiferroicity in *bi*-CrI$_3$/In$_2$Se$_3$ and *tri*-CrI$_3$/In$_2$Se$_3$ systems, in which the novel asymmetrical ME coupling, controlled by the switchable indirect orbital coupling, plays a critical role. In general, our findings here not only can overcome the two challenging problems (*i.e.*, problems i and ii) existing in the current 2D multiferroics, but also can be used to design other vdW multiferroics with different components and layer thickness.

*Acknowledgements*: B.H. and B.Y. thanks the helpful discussion with Drs. B. Cui and Y. Li. C.Y. and B.H. were supported by the Science Challenge Project (Grant No. TZ2016003) and NSAF U1930402. S.B.Z. was supported by U.S. DOE under Grant No. DE-SC0002623. All the calculations were performed at Tianhe2-JK at CSRC.

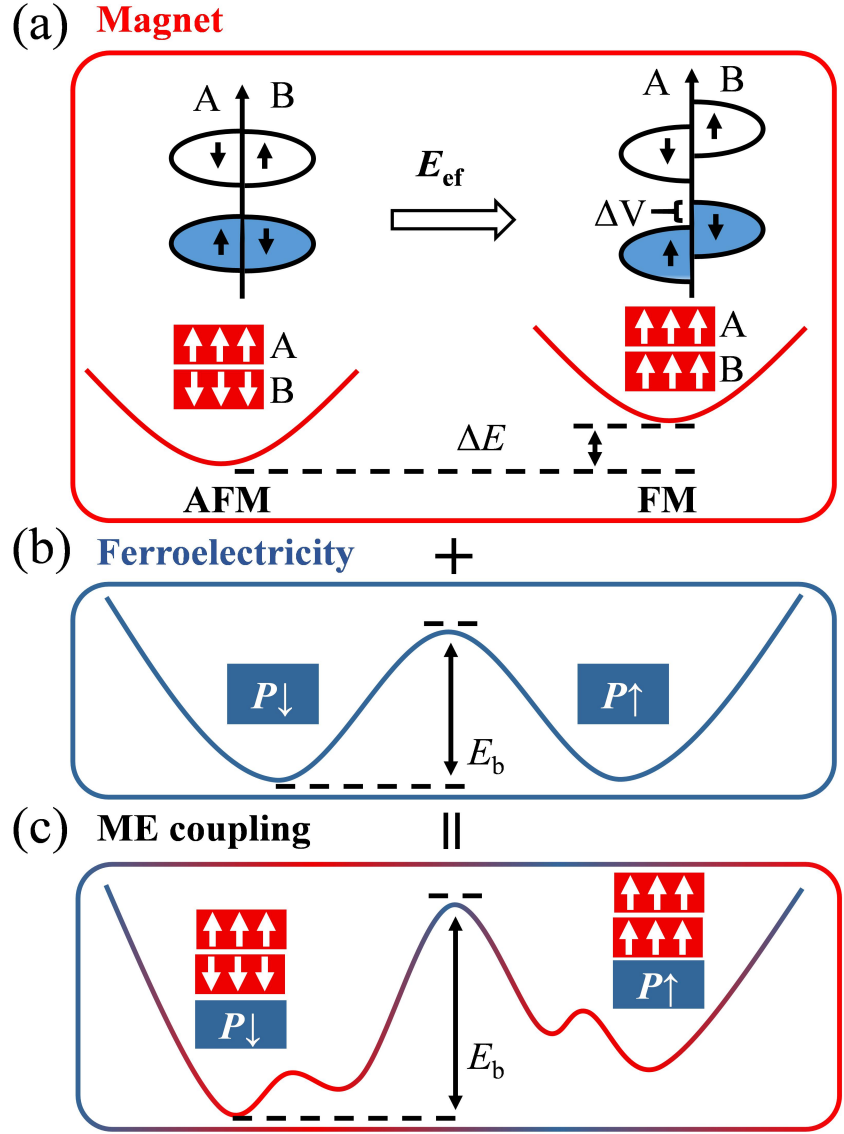

**FIG. 1.** Concept and schematic diagrams for realizing vdW multiferroics. (a) Electronic and magnetic states for a 2D bilayer magnet without (left panel) and with (right panel) an $E_{ef}$. Arrows denote spin orders in AB layers. (b) Ferroelectric double-well potential energy curve. $P\uparrow$ and $P\downarrow$ denote two opposite FE polarizations, while $E_b$ is the energy barrier between the $P\uparrow$ and $P\downarrow$ states. (c) Realization of a vdW multiferroics via an asymmetrical ME coupling between the bilayer magnet and ML FE systems.

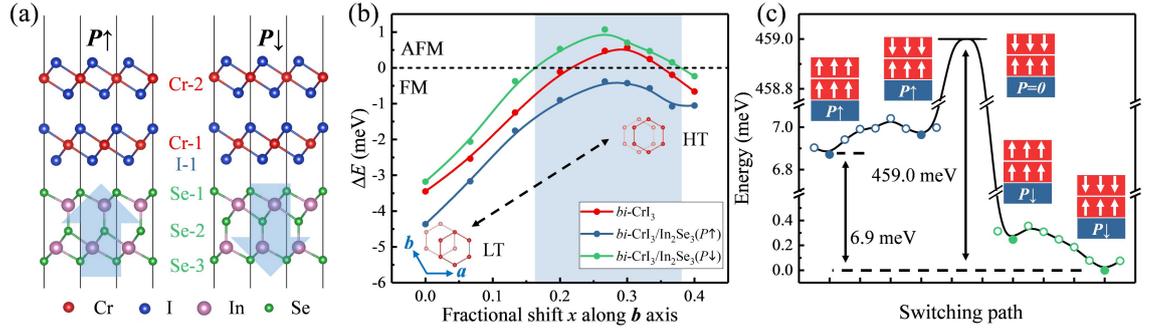

**FIG. 2.** Realization of vdW multiferroics in *bi*-CrI$_3$/In$_2$Se$_3$. (a) Side views of *bi*-CrI$_3$/In$_2$Se$_3$ with $P\uparrow$ (left panel) and $P\downarrow$ (right panel). (b) Energy difference ($\Delta E$) between the FM and AFM states in the free-standing *bi*-CrI$_3$ and *bi*-CrI$_3$/In$_2$Se$_3$, as a function of the fractional shift, which is obtained by sliding the top CrI$_3$ layer along the ***b*** axis with respect to the bottom layer. *x*=0.0 is the low-temperature (LT) phase, while *x*=1/3 is the high-temperature (HT) phase. The shaded area is where the FM-$P\uparrow$ and AFM-$P\downarrow$ states in Fig. 1(c) have been realized in *bi*-CrI$_3$/In$_2$Se$_3$. (c) Switching path between the FM-$P\uparrow$ and AFM-$P\downarrow$ states for the HT-phase *bi*-CrI$_3$/In$_2$Se$_3$. Blue and green circles are different magnetic states in the $P\uparrow$ and $P\downarrow$ configurations, respectively. When *P*=0, the system is non-ferroelectric.

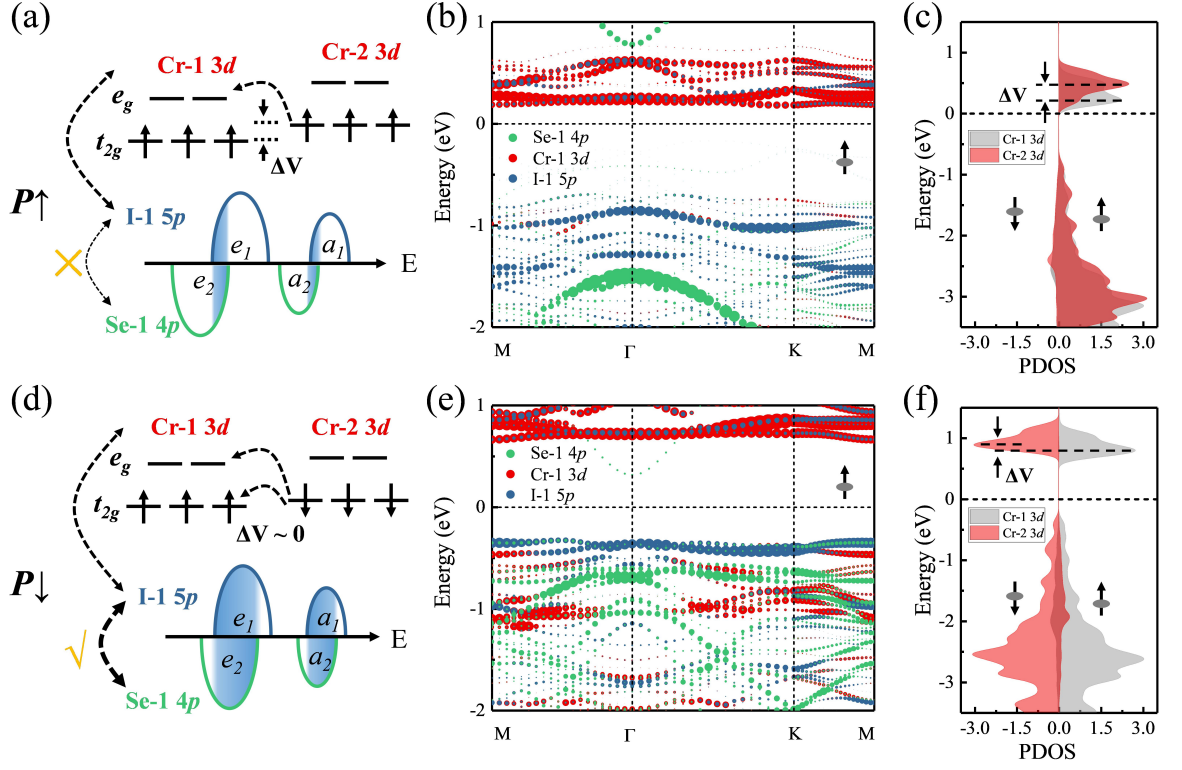

**FIG. 3.** Mechanism of multiferroicity in *bi*-CrI$_3$/In$_2$Se$_3$ where (a)-(c) are for *P*↑ and (d)-(f) for *P*↓. (a) and (d): Schematic diagrams of orbital coupling between interfacial I-1 and Se-1 *p* orbitals as well as the virtual orbital hopping path between empty Cr-1 $e_g$ and occupied Cr-2 $t_{2g}$ orbitals. Here, ΔV is the relative energy difference between the two $t_{2g}$ (and $e_g$) orbitals. In the lower part of the diagrams, $a_1$ ($a_2$) and $e_1$ ($e_2$) are the single and doubly-degenerate *p* orbitals of I-1 (Se-1) atoms, respectively. (b) and (e): Orbital-projected band structures (spin-up channel). (c) and (f): Projected densities of states (PDOS).

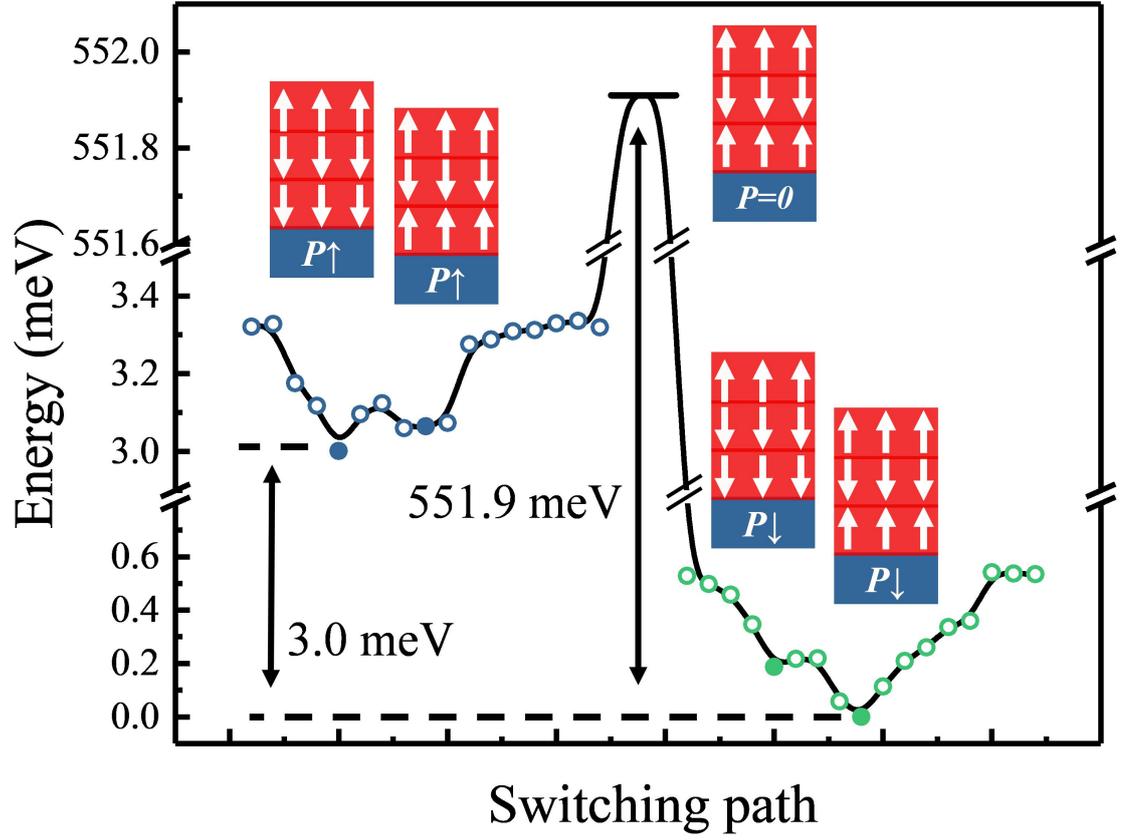

**FIG. 4.** Realization of vdW multiferroics in *tri*-CrI$_3$/In$_2$Se$_3$. Switching path of the HT-phase *tri*-CrI$_3$/In$_2$Se$_3$. Blue and green circles represent different magnetic states in the *P↑* and *P↓* configurations, respectively. *P* = 0 is the non-polarized configuration.